\documentclass[prl,aps,twocolumn]{revtex4}
\usepackage{epsfig,amsmath}
\usepackage{bm}
\def\Fbox#1{\vskip1ex\hbox to 8.5cm{\hfil\fboxsep0.3cm\fbox{%
\parbox{8.0cm}{#1}}\hfil}\vskip1ex\noindent}
\newcommand{\B}[1]{{\bm{#1}}}
\newcommand{\C}[1]{{\mathcal{#1}}}

\newcommand{\Onecol} {\begin{widetext} \onecolumngrid} 
\newcommand{\Twocol} {\end{widetext} \twocolumngrid} 

\newcommand{\be}{\begin{equation}}
\newcommand{\ba}{\begin{array}}
\newcommand{\bea}{\begin{eqnarray}}
\newcommand{\bfi}{\begin{figure}}
\newcommand{\ee}{\end{equation}}
\newcommand{\ea}{\end{array}}
\newcommand{\eea}{\end{eqnarray}}
\newcommand{\efi}{\end{figure}}

\def\Re{${\C R}\mkern-3.1mu e$} 
\def\RE{{\C R}\mkern-3.1mu e} 
\newcommand{\Sub}[1]{_{_{\text {#1}}}} 
\begin{document}
\title{Drag Reduction by a Linear Viscosity Profile}
\author{Elisabetta De Angelis$^{1,2}$, Carlo Casciola$^1$, Victor S.
L'vov$^2$, Anna Pomyalov$^2$, Itamar Procaccia$^2$ and Vasil
Tiberkevich$^2$} \affiliation{$^1$ Dip. Mecc. Aeron., Universit\`a
di Roma ``La Sapienza", Via Eudossiana 18, 00184, Roma, Italy\\
$^2$Dept. of Chemical Physics, The Weizmann Institute of Science,
Rehovot, 76100 Israel} 
\begin{abstract}
Drag reduction by polymers in turbulent flows raises an apparent
contradiction: the stretching of the polymers must increase the
viscosity, so why is the drag reduced? A recent theory proposed
that drag reduction in agreement with experiments is consistent
with the effective viscosity growing linearly with the distance
from the wall. With this self consistent solution the reduction in
the Reynolds stress overwhelms the increase in viscous drag. In
this Letter we show, using Direct Numerical Simulations, that a
linear viscosity profile indeed reduces the drag in agreement with
the theory and in close correspondence with direct simulations of
the FENE-P model at the same flow conditions.

\end{abstract}
\maketitle

The addition of few tens of parts per million (by weight) of
long-chain polymers to turbulent fluid flows in channels or pipes
can bring about a reduction of the friction drag by up to 80\%
\cite{49Toms,75Vir,97VSW,00SW}. In spite of a large amount of
experimental and simulational data, the fundamental mechanism has
remained under debate for a long time \cite{69Lu,90Ge,00SW}. Since
polymers tend to stretch in a turbulent flow, increasing thus the
bulk viscosity, it appears contradictory that they should reduce
the drag. There must exist a mechanism that compensates for the
increased viscosity. Indeed, drag is caused by two reasons, one
viscous, and the other inertial, related to the momentum flux from
the bulk to the wall. For a fixed rate (per unit mass) of momentum
generated by the pressure gradient, reducing the momentum flux can
reduce the drag. In a recent theory of drag reduction in wall
turbulence~\cite{03LPPT} it was proposed that the polymer
stretching gives rise to a self-consistent effective viscosity
that increases with the distance from the wall. Such a profile
reduces the Reynolds stress (i.e. the momentum flux to the wall)
more than it increases the viscous drag; the result is drag
reduction. The aim of this Letter is to substantiate this
mechanism for drag reduction on the basis of Direct Numerical
Simulations.

\begin{figure}
\centering \vskip -.25cm
\includegraphics[width=0.46\textwidth]{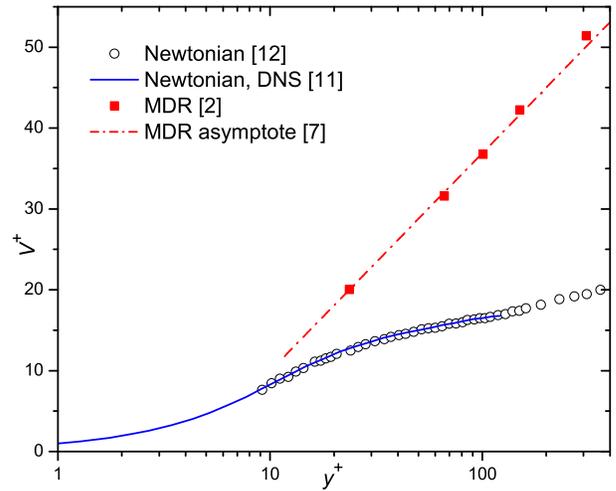}
\caption{Mean velocity profiles as a function of the distance from
the wall [in ``wall" units cf. Eq. (\ref{red})]. The solid line
(numerical simulations \cite{03ACLPP}) and the experimental points
(open circles) \cite{99WMH} represent the Newtonian results with
von-Karman's log law of the wall observed for $y^+> 20$. The red
data points (squares) \cite{75Vir} represent the MDR asymptote.
The dashed red curve represents the theoretical universal MDR
asymptote, Eq. (\ref{final})~\cite{03LPPT}.} \label{profiles}
\end{figure}

The onset of turbulence in channel or pipe flows increases the
drag dramatically. For Newtonian flows (in which the kinematic
viscosity is constant) the momentum flux is dominated by the
so-called Reynolds stress, leading to a logarithmic (von-Karman)
dependence of the mean velocity on the distance from the wall
\cite{00Pope}. However, with polymers, the drag reduction entails
a change in the von-Karman log law such that a much higher mean
velocity is achieved. In particular, for high concentrations of
polymers, a regime of maximum drag reduction is attained (the
``MDR asymptote"), independent of the chemical identity of the
polymer \cite{75Vir}, see Fig. \ref{profiles}.
In a recent theoretical paper \cite{03LPPT} the fundamental
mechanism for this phenomenon was elucidated: while momentum is
produced at a fixed rate by the forcing, polymer stretching
results in a suppression of the momentum flux from the bulk to the
wall. Accordingly the mean velocity in the channel must increase.
It was shown that when the concentration of the polymers is large
enough there exists a new logarithmic law for the mean velocity
with a slope that fits existing numerical and experimental data.
The law is universal, thus explaining the MDR asymptote.

To see how this mechanism works, consider the modified
Navier-Stokes (NS) equation for the polymer
solutions~\cite{87BCAH,94BE} :
\begin{equation}
{\partial \B U}/{\partial t} +\B U\cdot \B \nabla \B U=-\B \nabla
p +\B \nabla\cdot \B{{\cal T}} +\nu_0 \nabla^2 \B U \ , \label{FP}
\end{equation}
where $\nu_0$ is the kinematic viscosity of the carrier fluid and
$\B{{\cal T}}$ is the extra stress tensor that is due to the
polymer. Denoting the polymer end-to-end vector distance
(normalized by its equilibrium value) as $\B r$, the average
dimensionless extension tensor $ \B {{\cal R}}$ is ${\cal
R}_{ij}\equiv \langle r_i r_j \rangle$, and the extra stress
tensor is (with $\omega_{ij}\equiv
\partial U_i/\partial x_j$),
\begin{equation}\label{tau-polymer}
\B{{\cal T}} =\nu_p \left(\B\omega\cdot \B {{\cal R}}+ \B {{\cal
R}}\cdot \B\omega^{\scriptscriptstyle\rm T} - {\partial \B {{\cal
R}}}/{\partial t} -\B U\cdot \B \nabla \B {{\cal R}}\right) \ .
\end{equation}
Here $\nu_p$ (proportional to the polymer concentration) is the
polymeric contribution to the viscosity in the limit of zero
shear. Note that expression (\ref{tau-polymer}) for the extra
stress tensor is valid for any dumbbell model of the polymeric
molecule (i.e., FENE-P, Hookean, etc.).

In \cite{03LPPT}, two simplifying approximations led to a transparent
semi-quantitative theory of drag reduction. The first approximation
is to ignore the fluctuations of $\B {{\cal R}}$ as compared to its
mean and to take $ \B {{\cal R}}\approx \langle \B {{\cal R}} \rangle $ in
Eq.~(\ref{tau-polymer}). Then the main contribution of the stress
tensor $\B {{\cal T}}$ in the modified NS equation (\ref{FP}) can
be formally written in the form of some effective, $\B{{\cal
R}}$-dependent tensorial viscosity. The tensorial structure of the
effective viscosity describes different damping of different
components of the velocity fluctuations due to the preferred
orientation of polymer molecules along the mean shear. The second
approximation is to ignore the
preferential orientation and to replace ${\cal R}_{ij} \Rightarrow
{\cal R}\delta_{ij}$. These two approximations allow one to
simplify the dumbbell model (\ref{FP}) to a modified NS equation
with an effective viscosity, $\nu_0\Rightarrow \nu $, and
pressure, $p \Rightarrow P $:
\begin{equation}
\begin{array}{l}
{\partial U_i}/{\partial t} + U_j \nabla_j U_i=-\nabla_i P
+\nabla_j[\nu (\omega_{ij}+\omega_{ji})] \ , \\ \Big.
\nu=\nu_0+\nu_p {{\cal R}}\ , \quad P=p+{\partial \nu} /{\partial
t} +\B U\cdot \B \nabla \nu \ .
\end{array}\label{FPnew}
\end{equation}
Obviously, the polymer elongation, $\cal R$, depends on the
distance from the wall, leading to the corresponding dependence of
the effective viscosity. Notice that the above approximations
are uncontrolled. Their verification is one of the main goals of
this Letter. We demonstrate that the dynamical model Eq.
(\ref{FPnew}) contains the essential properties of the full FENE-P
model Eqs. (\ref{FP}), (\ref{tau-polymer}).

In Ref. \cite{03LPPT} we modified the Reynolds closure approach to
the situation in which the scalar viscosity cannot be neglected.
This approach, that was justified by considering a reasonable
model of the coil-stretch transition of the polymers, resulted in
the analytical form of the universal MDR logarithmic profile 
\begin{equation}
V^+(y^+) = \frac{1}{\kappa\Sub V}\ln\left(e\, \kappa\Sub V\,
y^+\right) \,,\quad \kappa\Sub V \approx 0.09 \ ,\label{final}
\end{equation}
presented in the wall units
\begin{equation}
\RE_\tau \equiv {L\sqrt{\mathstrut p' L}}/{\nu_0}\ , \ y^+ \equiv
{y \RE_\tau }/{L} \ , \ V^+ \equiv {V}/{\sqrt{\mathstrut p'L}} \ .
\label{red}
\end{equation}
Here $p'\equiv -\partial p/\partial x$ is fixed pressure gradients
in the streamwise direction $x$, $L$ is the half-width of the
channel (in the wall-normal direction $y$), and $\RE_\tau$ is the
friction Reynolds number. Equation~(\ref{final}) is in excellent
agreement with the MDR asymptote, see Fig.1. Another conclusion of
\cite{03LPPT} is that in the MDR regime the polymer extension
${\cal R}(y)$ is self-adjusted in order to provide a universal
linear profile of the effective viscosity 
\begin{equation}\label{nu-mdr}
\nu\Sub{MDR}(y) = \kappa\Sub V\sqrt{p'L}\ y \ .\end{equation}
Needless to say, in the viscous sublayer the viscosity is
Newtonian: $\nu=\nu_0$. It should be noted that the possibility of
drag {\em reduction} by {\em increasing} of the viscosity looks
somewhat paradoxical: in the usual Newtonian case with constant
viscosity, the drag is monotonically increasing with the
viscosity. The point is that for the polymer solutions the
effective viscosity is not constant anymore, it increases linearly
with the distance from the wall according to Eq.
(\ref{nu-mdr}). This point is the essential difference of the
theory \cite{03LPPT} from all previous ``viscous'' theories of
drag reduction (see, e.g., \cite{69Lu}). Therefore a crucial test
of the theory~\cite{03LPPT} is to introduce such a {\em linear}
viscosity profile to the NS Eq. (\ref{FPnew}) by hand, and see
whether we observe drag reduction together with its various
statistical aspects.

To this aim we simulate the effective NS Eq. (\ref{FPnew}) with
proper viscosity profiles (discussed below) and show that the
results are in semi-quantitative agreement with the corresponding
full FENE-P DNS. All simulations were done in a domain $2\pi
L\times2L\times 1.2\pi L$, with periodic boundary conditions in
the streamwise and spanwise directions, and with no slip
conditions on the walls that were separated by 2$L$ in the
wall-normal direction. An imposed mass flux and the same Newtonian
initial conditions were used. The Reynolds number \Re~(computed
with the centerline velocity) was $6000$ in all the runs. The grid
resolution is $96\times 129 \times96$ for the linear viscosity
profile runs and $96\times 193 \times96$ for the FENE-P run. The
latter was done with $De_\tau=52.7$, $\eta_p=0.1$, ${\cal
R}^2_{max}=1000$. For a definition of these parameters and
details of the numerical procedure see Ref.
\cite{02ACP}.

\begin{figure}
\includegraphics[width=0.45\textwidth]{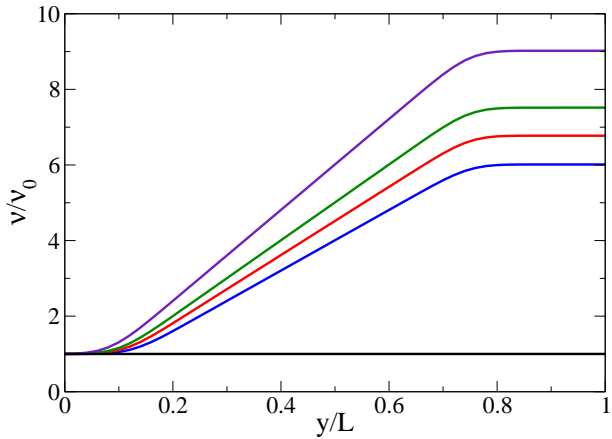}
\caption{The Newtonian viscosity profile and four examples of
close to linear viscosity profiles employed in the numerical
simulations. Solid black line: run N, solid blue: run R, red: run
S, green: T, violet: run U. } \label{visprof}
\end{figure}

The $y$ dependence of the scalar effective viscosity was close to
being piece-wise linear along the channel height, namely
$\nu=\nu_0$ for $y\le y_1$, a linear portion with a prescribed
slope for $y_1<y\le y_2$, and again a constant value for
$y_2<y<L$.
For numerical stability this profile was smoothed
out according to the differential equation
$$\displaystyle \frac{d^2 \nu}{d y^2} = \frac{C\nu_0}{\sqrt{2\pi}
\sigma L}\Big\{\exp
\Big[-\frac{(y-y_1)^2}{2
\sigma^2}\Big]-\exp\Big[-\frac{(y-y_2)^2}{2
\sigma^2}\Big]\Big\}\,,
$$
integrated with initial conditions $\nu(0)=\nu_0$, and $\nu^\prime
(L)=0$. We chose $\sigma=0.04L , \ y_1=L/C, \ y_2=3L/4$, while $C$
is the dimensionless value of the slope. Examples of four such
profiles are shown in Fig. \ref{visprof}. The slope of the linear
part of the viscosity profile [i.e., parameter $\kappa\Sub V$ in
(\ref{nu-mdr})] was varied for different runs and was smaller than
the theoretically predicted value for the MDR regime. The
simulations with $\nu_{\scriptscriptstyle\rm MDR}(y)$ require
larger \Re\ numbers to sustain the turbulence. Included in the
figure is the flat viscosity profile of the standard Newtonian
flow.
Since we keep the throughput constant, the runs differ in the
values of friction Reynolds numbers $\RE_\tau\equiv
\sqrt{\tau_w}L/\nu_0$ where $\tau_w$ is the average friction at
the wall $\tau_w\equiv \nu_0\, d U/dy$. The decreased value of
Re$_\tau$ is a manifestation of the drag reduction measured in
percentage as $DR\%=(\tau_w^{\scriptscriptstyle\rm
N}-\tau_w^{\scriptscriptstyle\rm E})/\tau_w^{\scriptscriptstyle\rm
N}$. The normalized slopes, the value of $\RE_\tau$ and the
percentage of drag reduction for these runs are summarized in
Table I.

\begin{table} [b]
\caption{DNS parameters for effective viscosity runs.}

\begin{center}
\begin{tabular}{||c|c|c|c|c||}
\hline \hline ~Case~ & $~C~$ &
$~Re_{\tau}~$ & $~DR\%~$ & $\kappa\Sub V$\\
\hline
N & 0 & 245 & -- & -- \\
R & 8 & 227 & 13.8 &~0.035~\\
S & 9 & 214 & 21.6 &~0.042\\
T & 10 & 197 & 36.9 &~0.051 \\
U & 12 & 185 & 42.0 &~0.065 \\
\hline\hline
\end{tabular}
\end{center}
\label{Table}

\end{table}

\begin{figure}
\centering
\includegraphics[width=0.45\textwidth]{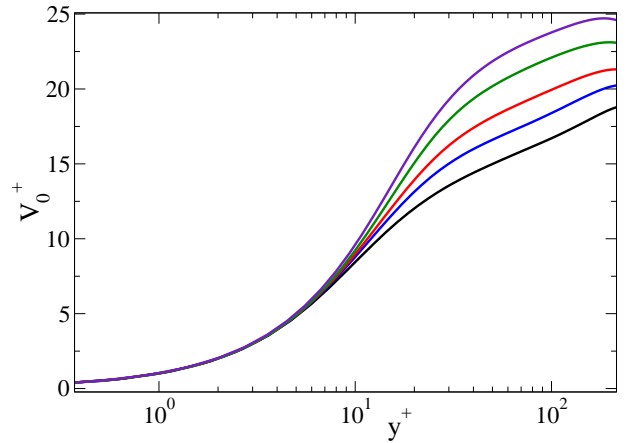}
\caption{The reduced mean velocity as a function of the reduced
distance from the wall. The line types correspond to those used in
Fig. \ref{visprof}.} \label{Vprofiles}
\end{figure}

In Fig. \ref{Vprofiles} we show the resulting profiles of
$V^+_0(y)$ vs. $y^+$. The line types are chosen to correspond to
those used in Fig. \ref{visprof}. The {\em decrease} of the drag
with the {\em increase} of the slope of the viscosity profiles is
obvious. Since the slopes of the viscosity profiles are smaller
than needed to achieve the MDR asymptote for the corresponding
$\RE_\tau$, the drag reduction occurs only in the near-wall region
and the Newtonian plugs are clearly visible.

It is most interesting to compare the effect of the linear
viscosity profile to the simulation of the FENE-P model in which
both the throughput {\em and} $\RE_\tau$ are the same. Such a
comparison was performed for the ``S'' viscous run, for which
$\RE_\tau=214$ and the FENE-P run with $\RE_{\tau}=212.5$.
\begin{figure}
\centering
\includegraphics[width=0.45\textwidth]{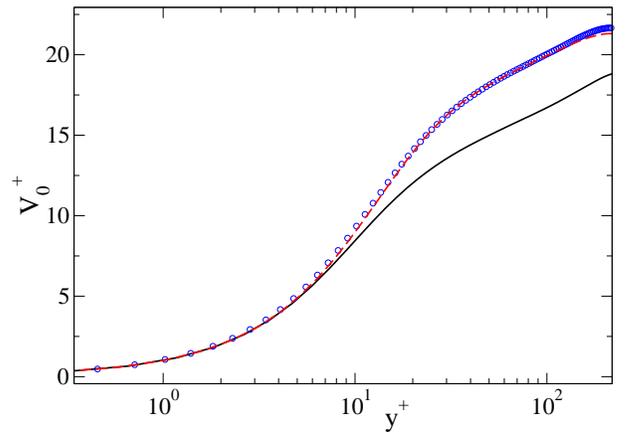}\\ ~\\ ~\\
\includegraphics[width=0.45\textwidth]{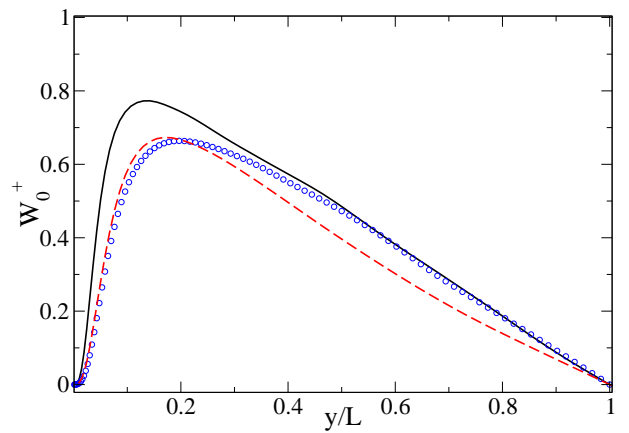}
\caption{Upper panel: the reduced mean velocity as a function of
the reduced distance from the wall. Lower panel: Reynolds stresses
across the channel. Continuous line: Newtonian. Dashed line:
linear viscosity profile. Symbols: FENE-P. } \label{compFENEP}
\end{figure}
\begin{figure}
\centering
\includegraphics[width=0.43\textwidth]{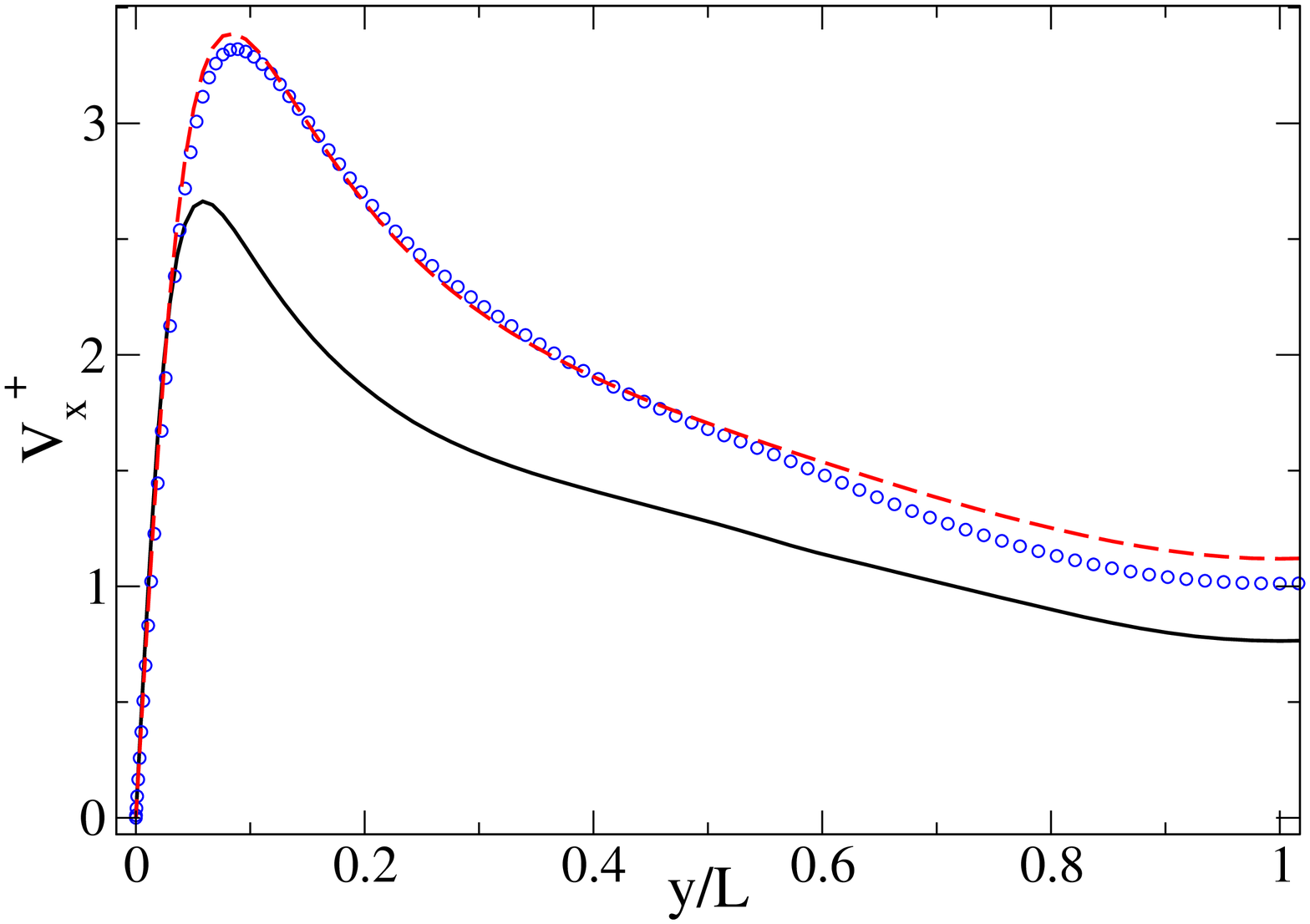} \vskip 1cm
\includegraphics[width=0.43\textwidth]{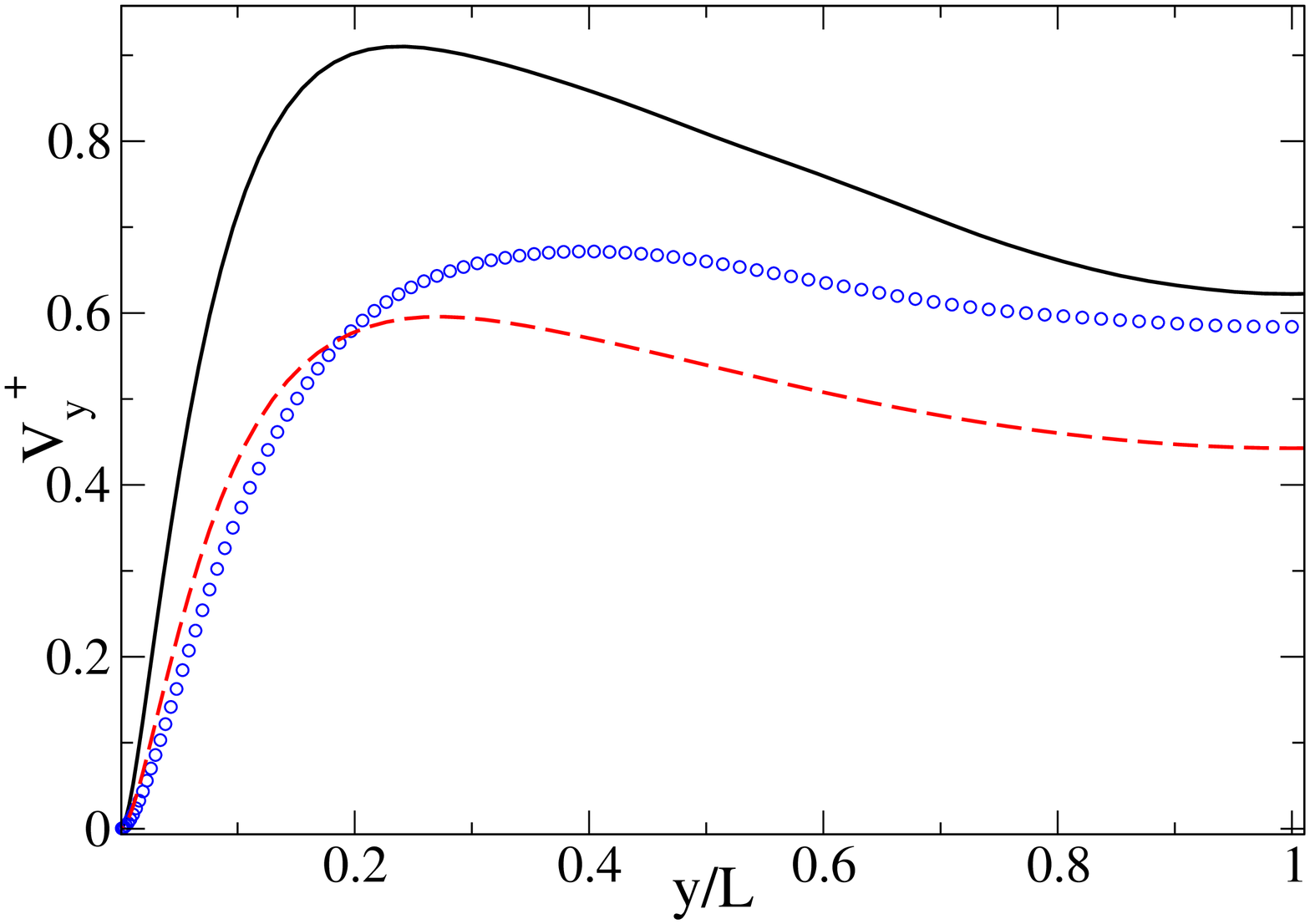}
\caption{The rms streamwise ( upper panel) and wall-normal (lower
panel) velocity fluctuations across the channel. The line types
correspond to those used in Fig. \ref{compFENEP}}
\label{compFENEP-rms}
\end{figure}
The results are presented in the two panels of Fig.
\ref{compFENEP}. In the upper panel the mean velocity profiles
(symbols for FENE-P and dashed line for the linear viscosity
profile) are seen to correspond very closely. The region with
increased slope of the mean velocity (e.g., the region where the
drag reduction occurs) is limited to approximately $y^+\le50$, or,
in natural units, to $y/L\le0.25$. In the outer region (Newtonian
plugs, $y^+\ge50$ or $y/L\ge0.25$) the polymer molecules in the
FENE-P model are supposedly unstretched and the polymeric
contribution to the effective viscosity is small. In contrast, in
the viscous model the effective viscosity is maximum in this
region and may be much larger than $\nu_0$.
Therefore, if the mechanism of the drag reduction in viscous model
is the same as in full FENE-P model, we expect that all
statistical quantities qualitatively coincide for both models in
the elastic region $y/L\le0.25$, but may differ in the Newtonian
plugs $y/L\ge0.25$.

In Fig. \ref{compFENEP}, lower panel, we show the normalized
Reynolds stresses. Clearly, in the elastic region $y/L<0.25$ both
drag-reducing models coincide; this is a strong evidence that the
reduced model (\ref{FPnew}) captures all essential properties of
the full model (\ref{FP}) and the mechanisms of drag reduction are
the same in both cases.

Another important characteristic of drag-reducing flows is
behavior of the root-mean-square (r.m.s.) velocity fluctuations.
The increase in r.m.s. streamwise ($V^+_x$) and decrease in r.m.s.
wall-normal ($V^+_y$) velocity fluctuations were observed in many
experiments and numerical simulations of drag-reducing flows. In
Fig. \ref{compFENEP-rms} we compare these quantities for FENE-P
and ``S'' runs.
Clearly, the correct trend of the r.m.s velocity fluctuations is
observed, indicating that the important features of the mechanism
of drag reduction are correctly captured by the model. An almost
quantitative agreement is reached in the region where drag reduction
actually occurs ( $y/L \approx 0.1-0.3$).

In conclusion, we showed that the simple linear viscosity model
(\ref{FPnew}), (\ref{nu-mdr}) faithfully demonstrates
drag-reducing properties and, surprisingly enough, the amount of
drag reduction increases with the increase of the slope of the
viscosity profile. Even more interestingly, the behavior of
objects like the Reynolds stress or the velocity fluctuations in the
elastic sublayer are in close correspondence with the full FENE-P
model, indicating that the mechanisms of drag reduction proposed
in \cite{03LPPT} operates similarly in both cases.

This work was supported in part by the US-Israel BSF, The ISF
administered by the Israeli Academy of Science, the European
Commission under a TMR grant and the Minerva Foundation, Munich,
Germany.

\end{document}